\documentclass[12pt,a4paper]{article}
\pdfoutput=1

\usepackage{slashed,braket}
\usepackage{jheppub}

\usepackage{tikz}
\usetikzlibrary{arrows,shapes,calc}
\usetikzlibrary{decorations.pathreplacing} 
\usetikzlibrary{decorations.pathmorphing}
\usetikzlibrary{decorations.markings}
\tikzset{
	 >=stealth',
    v/.style={decorate, decoration={snake}, draw,thick},
	provector/.style={decorate, decoration={snake,amplitude=2.5pt}, draw},
	antivector/.style={decorate, decoration={snake,amplitude=-2.5pt}, draw},
    f/.style={draw=black, postaction={decorate},
        decoration={markings,mark=at position .55 with {\arrow[draw=black,very thick]{>}}}},
    fb/.style={draw=black, postaction={decorate},
        decoration={markings,mark=at position .55 with {\arrow[draw=black,very thick]{<}}}},
}

\title{Angular distributions as lifetime probes}

\author{Jeff Asaf Dror}
\emailAdd{ajd268@cornell.edu} 
\author{and Yuval Grossman}
\emailAdd{yg73@cornell.edu} 
\affiliation{Department of Physics, LEPP, Cornell University, Ithaca, NY 14853 \vspace*{8pt}}

\abstract{
If new TeV scale particles are discovered, it will be important to determine their width. There is, however, a problematic region, where the width is too small to be determined directly, and too large to generate a secondary vertex. For a collection of colored, spin polarized particles, hadronization depolarizes the particles prior to their decay. The amount of depolarization can be used to probe the lifetime in the problematic region. In this paper we apply this method to a realistic scenario of a top-like particle that can be produced at the LHC. We study how depolarization affects the angular distributions of the decay products and derive an equation for the distributions that is sensitive to the lifetime.
}

 \arxivnumber{}

\begin{document} 
\maketitle 
\flushbottom
\section{Introduction}
Measuring the lifetime of new particles is a powerful method to probe Beyond Standard Model (BSM) theories. While the first run at the LHC did not yield any BSM particles, we still hope to find such states. If new particles are found with short lifetimes or equivalently large widths, $\Gamma \gtrsim 1\, \mbox{GeV}$, then their widths can be measured directly. If these particles live much longer, $\Gamma \lesssim 10^{-4} \,\mbox{eV} $, then we can measure their lifetimes using a displaced vertex. However, there is currently no technique to measure particle lifetimes in the ``problematic region,'' 
\begin{equation}
  10^{-4} \,\mbox{eV} \lesssim  \Gamma \lesssim 1 \,\mbox{GeV} \,. \label{eq:lifetime}
\end{equation} 
This may not be an issue as many BSM theories do not predict any particles with lifetimes in this region. Nevertheless, there are many well motivated BSM theories with particles in the problematic region, such as, $Z'$-mediated supersymmetry (SUSY) breaking models (e.g. the wino next-to-lightest supersymmetric particle (NLSP))~\cite{Yavin2008}, split SUSY (e.g the neutralino with a large neutralino mass)~\cite{Romanino2005}, Gauge-mediated broken SUSY (e.g. the NLSP with breaking scale $\lesssim 10^{6}\,\mbox{GeV}$)~\cite{ArkaniHamed2005}, SUSY with a heavy scalar mass scale (e.g. gluino with heavy scalar mass $ \lesssim 10^{4} \,\mbox{TeV} $)~\cite{Tobioka2012}, Dynamic R-parity violating supersymmetric models (e.g. the gluino as the lightest supersymmetric particle with large gluino mass)~\cite{Csaki2013}, minimal flavor violation SUSY models (e.g. the stop except for very small values of $ \tan \beta $)~\cite{Csaki2012,Smith2008}, and GUTS in warped extra dimensions(e.g. next-to-lightest $Z_3$ charged particles)~\cite{Agashe2005}.

The problem of how to probe lifetimes in the problematic region was discussed in detail in ref.~\cite{Grossman2008}, where a new technique to measure the lifetime of a hypothetical heavy colored particle was suggested. The article focused on a particle with the same quantum numbers as the top, but its mass and width were kept as unknowns. In ref.~\cite{Grossman2008}, however, the polarization of the top-like particle was assumed to be known. There was no discussion of how such a measurement can be done, nor about the interplay between the measurement and the lifetimes. In our current work we apply the principles of ref.~\cite{Grossman2008} and present a calculation of a quantity that can be measured at the LHC and future colliders. Furthermore, we present the theoretical difficulties involved in choosing a suitable channel to measure the particle lifetime and discuss the importance of using a convenient spin basis. 

We briefly discuss the general principles of the method here and leave the details to ref.~\cite{Grossman2008}. Suppose a top-like particle is produced with spin up (note that we will often refer to our hypothetical particle as a top or $t$ quark while we really mean a top-like quark with a longer lifetime than the top). After a short time, of order $1/\Lambda_{QCD}$, the colored particle forms a hadron through hadronization, with mesons forming about $90\%$ of the time~\cite{Chliapnikov1999} (in our discussion we focus on meson production and decay, though adding in baryons is straightforward~\cite{Grossman2008}). Since the hyperfine splitting is small compared to $\Lambda_{QCD}$, after hadronization the system is in its QCD ground state without resolving the hyperfine splitting. That is, it produces an incoherent mix,
\begin{equation} 
50 \% \,\ket{++} \,\oplus 50\, \% \,\ket{+-}  \, ,\label{eq:mesonmix}
\end{equation} 
where $\oplus$ denote an incoherent sum and the state $\ket{++}$ ($\ket{+-}$) denotes a meson with top spin up and the light degree of freedom with spin up (down). Excluding decays, the mass eigenstates of the system are given by the triplet and singlet state which we denote by $T(s, m_s)$ ($s=0,1$ is the spin and $m_s$ is the spin along the quantization axis). These states are related to the earlier ones by
\begin{equation} 
\ket{++} = T(1,1), \qquad \ket{+-} = \frac{T(1,0)+T(0,0)}{\sqrt{2}} \,.\label{eq:states}
\end{equation} 
One source of depolarization is due to the triplet state decaying into the singlet with a lifetime that we denote by $1/\Delta \Gamma$. Another source is due to the top that is initially in the $\ket{+-}$ state, which oscillates between a spin up and spin down with a timescale of $1/\Delta m$ where $\Delta m$ is the mass splitting between the triplet and singlet states. The third timescale involved is the weak decay width of the top quark, $1/\Gamma$. The calculation of $\Delta m$ and $\Delta \Gamma$ is discussed in ref.~\cite{Grossman2008}, and for our discussion we assume them to be known. Then, the idea is to measure an observable that is sensitive to the polarization of the top quarks. The amount of depolarization gives us the sensitivity to $\Gamma$. 

Before we go on, we emphaise the basic assumptions that we use (see also ref.~\cite{Peskin1994,Falk1996,Mehen1996}). Our treatment is valid as long as all the relevant scales, that is, $\Gamma$, $\Delta m$ and $\Delta\Gamma$ are much smaller than $\Lambda_{QCD}$. Under this assumption, when studying spin-flipping interactions, we only need to consider the ground state (the state with orbital angular momentum, $L = 0$, and principle quantum number, $n = 0$). This is because any excited states have energies of order $\Lambda_{QCD}$ or more above the ground state and so any excited mesons would quickly decay to the ground state with a rate ${\cal O}(\Lambda_{QCD})$.

\begin{figure} 
\center
\includegraphics[width=12cm]{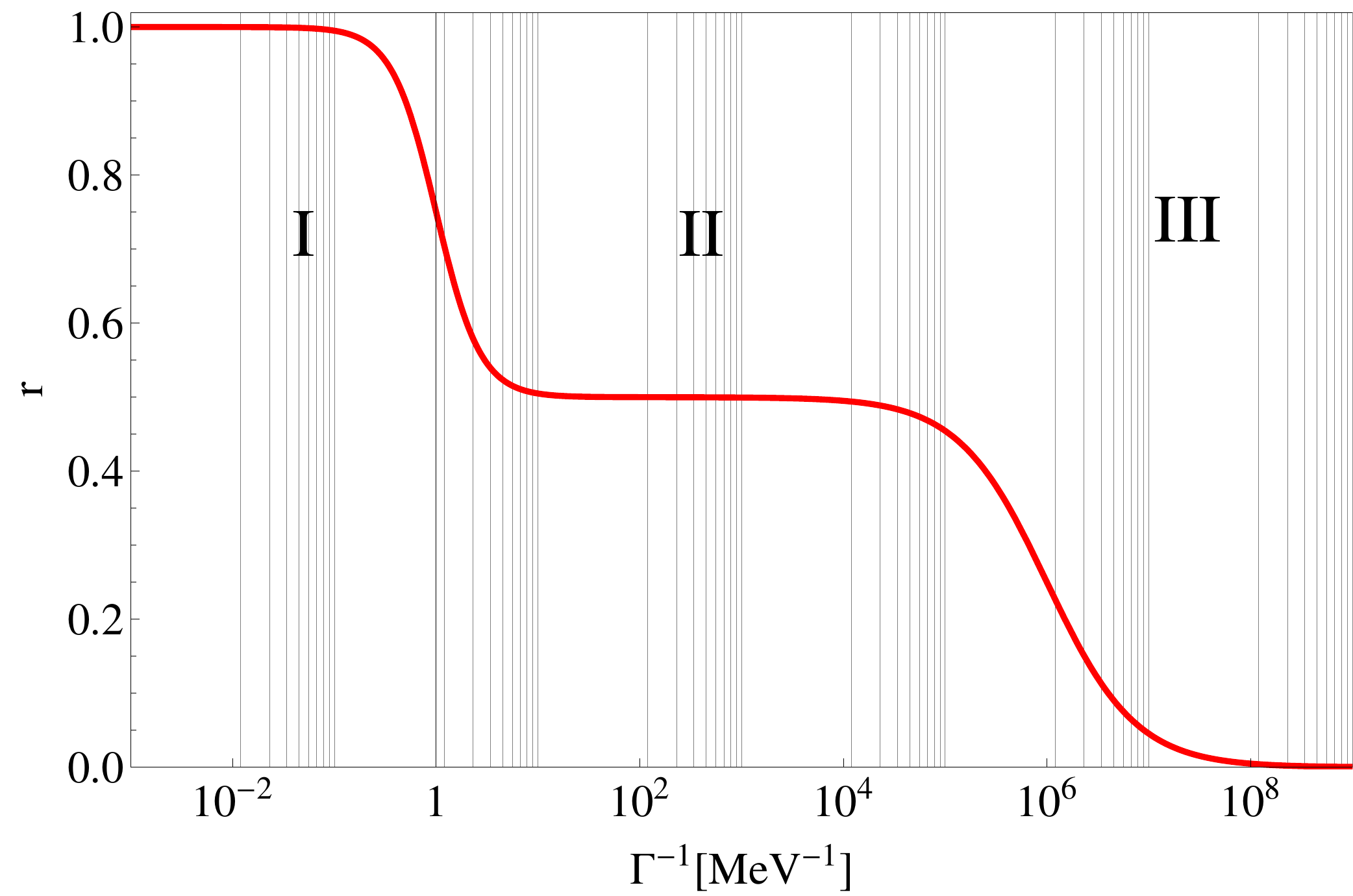}
\caption{$r$ as a function of the lifetime for $\Delta \Gamma = 1 \,\mbox{eV} $ and $ \Delta m = 1\, \mbox{MeV}$. Particles with lifetimes in regions I , II, and III have different levels of depolarization prior to decay. By measuring $r$ we have a direct measurement of the lifetime.}\label{fig:r}
\end{figure}

We use as our observable the spin of the top projected onto a quantization axis. The average spin in this direction of a collection of tops as a function of time, excluding top quark decay, is given by 
\begin{equation} 
{\rm pol}(t)\equiv \frac{ \left\langle s  \right\rangle (t) }{\left\langle s \right\rangle (0)} = \frac{1}{2} \left( e ^{-\Delta \Gamma t} + \cos \Delta m t \right). \label{eq:sz}
\end{equation} 
Note that this is the spin projection for the top-like quarks, not the mesons themselves. We have access to the spin of the tops through their decay products. Since we do not have good enough time resolution to measure ${\rm pol}(t)$ directly, what we measure is the time integrated value multiplied by the exponential probability density function:
\begin{equation}
r  \equiv \int d t \Gamma e ^{ - \Gamma t} \frac{ \left\langle s \right\rangle (t) }{ \left\langle s \right\rangle (0) } =
 \frac{1}{2} \left( \frac{1}{ 1 + x ^2 } + \frac{1}{ 1 + y } \right) \,, \label{eq:r}
\end{equation}
where we defined,
\begin{equation} 
x \equiv \frac{\Delta m}{\Gamma} , \qquad 
y \equiv \frac{\Delta \Gamma}{\Gamma}. \label{eq:xy}
\end{equation}
(note that ref.~\cite{Grossman2008}, contains an error and defines $y \equiv \Delta \Gamma / 2\Gamma $ while it should be defined as the above.) A plot of $r$ as a function of $\Gamma$ is shown in Fig.~\ref{fig:r}. In a hypothetical world where $\Delta m$, $\Delta \Gamma$, and $r$ are known to infinite precision, $\Gamma$ can be extracted. In practice, however, we can only split the problematic region into three. In region I, where the lifetime is short, the decaying particles do not feel the hadronization, as is the case for the real top quark~\cite{Bigi1986}. In region II, at the time its polarization is measured by the weak decay, the top has lost about half its initial polarization due to oscillation. In region III, the top has lost all its polarization prior to its decay due to both oscillations and to decay from the triplet to the singlet state. Because of the staircase structure of $r(\Gamma)$, the uncertainties of our input parameters, $\Delta m$ and $\Delta \Gamma$, typically have a very small impact on the results. We note that our method is similar to discussions regarding heavy quark meson oscillations, see for example ref.~\cite{Gay2001,Chavez2012}.

At this point all the basic physics is set, the big question left is how to practically measure this effect. We cannot create exotic particles, trap them, polarize them, and measure their angular momentum. Instead we propose a method to make use of current detectors available at the LHC. One potential concern in this study is that we are completely reliant on our ability to produce polarized particles. While not simple, this turns out to be straightforward and has been shown to be feasible at the LHC in the SM in both single top production~\cite{ATLAS2013} and ditop channels~\cite{Parke1996} due to its chiral nature. A second challenge is measuring the polarization itself. The spin of each particle is unattainable at the LHC, but the angular distributions can be measured. For a review of such methods, see, for example,~\cite{Yavin2007,Tattersall2011,Singh2009}. For a chiral theory the angles of the decay products depend on the polarization of the parent particle. Thus depolarization effects will appear in the angular distributions, and in particular, a forward backward asymmetry. This, in turn, will enable us to probe the lifetime. Our idea is therefore, to define a forward backwards asymmetry that changes as a function of $r$ and thus can be used to extract the lifetime.

In this work our emphasis is on studying a working example for the lifetime measurement. While depending on the specific particle production and decay channel the details will change, the ideas will should be the same. Furthermore, we do not attempt a full study with semi-realistic detector simulation but leave such investigation for future studies. Furthermore, we focus only on a top-like quark which has a simple ground state spectrum. More complicated situations, for example, a gluino or KK-gluon, can be treated in a similar way. Yet, we do not consider such cases at all in this work.

\section{General Formalism}
\begin{figure} \center
\begin{tikzpicture}[scale=1.5]
\draw [thick] (0,4.5) to [out=180,in=0] (-0.5,3) to [out=0,in=180] (0,1.5); \node[left] at (-0.5,3) {$ \substack{\mbox{initial} \\ \mbox{state}} $};
\draw[f,ultra thick] (0,4) -- (1,3.5);
\draw[f,ultra thick] (0,2) -- (1,2.5);

\draw [ultra thick,line width=2.5pt,line cap=round, dash pattern=on 0pt off 2\pgflinewidth,black] (5-0.06,2.5-0.05-0.05) to [out=25,in=25] (5-0.06,4.5);

\draw[f, ultra thick,red] (2,3.25) -- (3.66,3.416); \draw[f, ultra thick,blue] (3.66,3.416) -- (5.7,3.62);  \draw[f, ultra thick,green]  (5.7,3.62) -- (7,3.75); \draw[ultra thick, ->,red] (3,3.35)  -- (3.75,4.1); \node[above] at (3.75,4.1) {\textcolor{red}{$ s _t $}}; \draw [thick,<-] (3.3,4) to [out=-135,in=135] (3.25,3.6) to [out=-45,in=-135] (3.6,3.6); \draw[red,fill=white,ultra thick] (3,3.35) circle [radius=.3]; \node at (3,3.35) {\large $ t $} ;

\draw[f, ultra thick] (2,3) -- (3,2.5);
\draw[f, ultra thick] (2,2.75) -- (3,2.25);
\draw[f, ultra thick] (2,2.5) -- (3,2);

\draw[ultra thick, ->,blue] (5.075-0.06,4.5)  -- (5.825-0.06,5.25); \node[above] at (5.825-0.06,5.25) {\textcolor{blue}{$ s _t $}}; \draw [thick,<-] (5.4-0.06,5.2) to [out=-135,in=135] (5.4-0.06,4.8) to [out=-45, in=-135] (5.8-0.06,4.8);
\draw[blue, fill=white,ultra thick] (5.075-0.06,4.5) circle [radius=.2];
\node[black] at (5.075-0.06,4.5) {$ u $} ; 

\draw[ultra thick, ->,blue] (5.075-0.06,3.5)  -- (5.825-0.06,4.25); \node[above] at (5.825-0.06,4.25) {\textcolor{blue}{$ s _t $}}; \draw [thick,->] (5.4-0.06,4.2) to [out=-135,in=135] (5.4-0.06,3.8) to [out=-45, in=-135] (5.8-0.06,3.8);
\draw[blue, fill=white,ultra thick] (5-0.06,3.5) circle [radius=.3]; \node at (5-0.06,3.5) {\large $ t $};
\draw [ultra thick,line width=2.5pt, line cap=round, dash pattern=on 0pt off 2\pgflinewidth,black] (5-0.06,2.5-0.05-0.05) to [out=225-25,in=210] (5-0.06,4.5); 

\draw[f,ultra thick,green] (7,3.75) -- (8,3); \node[below] at (8.,3) {$ b$ } ;
\draw[v,ultra thick,green] (7,3.75) -- (8,4.5); \node[above] at (8,4.5) {$ W ^+ $ } ;

\draw [thick] (3,2.75) to [out=0,in=180] (3.25,2.25) to [out=180,in=0] (3,1.75);
\node[right] at (3.1,2.) {$ \substack{\mbox{extra} \\ \mbox{outgoing} \\ \mbox{particles}}$ };e

\draw[thick,->](0,1) -- (8,1) ;
\node[below] at (8,1) {time};

\draw[fill=gray,ultra thick] (1.5,3) circle [radius=.75];
\end{tikzpicture}
\caption{(color online). A general process considered here. Two particles combine producing a top-like quark [red], which hadronizes [blue] and subsequently decays [green] after a time $t$. A $ \bar{ u} $ quark is chosen as the light degree of freedom, though it can be any light quark. $ s _t $ denotes our chosen spin quantization axis. The interaction between the spin of the  $ \bar{u} $ and $t$ may provide a spin flip in the top as shown.}
\label{fig:genprocess}
\end{figure}
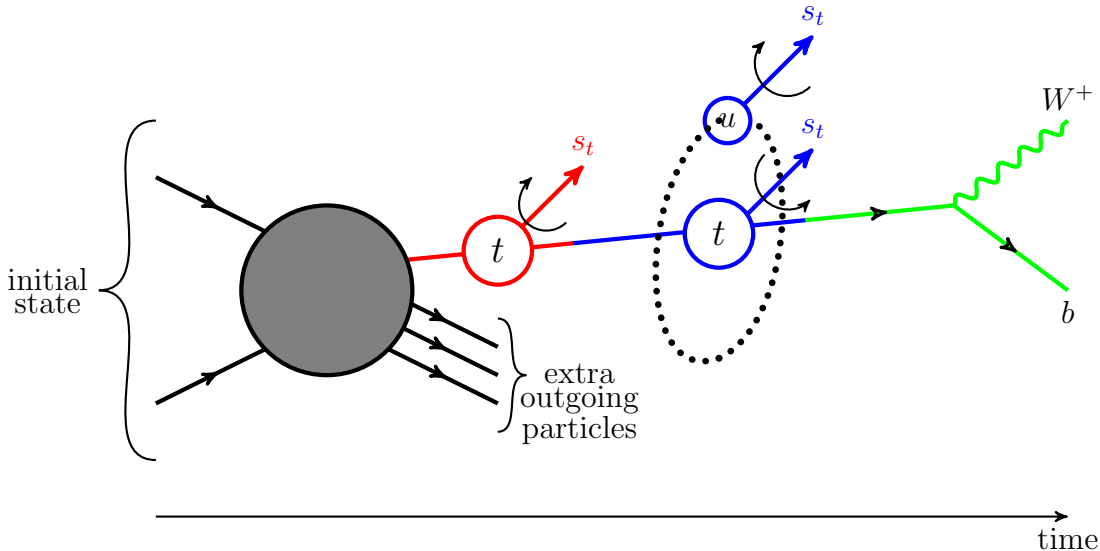
A general process is shown in Fig.~\ref{fig:genprocess} in which a top is produced, undergoes hadronization, and then decays. We first outline the calculation without hadronization and then in the non-perturbative effects. We assume the width of the top is much less than its mass (as is the case for any particle for which this method is useful) and work far from the top production threshold. We can now write the square of the matrix element for production and decay of a single top quark in an arbitrary channel as (for a discussion on these methods with regards to the top quark see e.g.~\cite{Singh2006}),
\begin{equation} 
\left| {\cal M} \right| ^2 = \frac{ \pi \delta \left( t ^2 - m ^2 \right) }{ m\Gamma }\sum _{ \lambda  , \lambda '} \rho  \left( \lambda,  \lambda ' \right) \Gamma \left( \lambda , \lambda ' \right)\,, \label{eq:rhogam}
\end{equation} 
where $ \lambda$ and $ \lambda'$ label the spin of the top quark and 
\begin{equation}
 \rho \left( \lambda , \lambda ' \right) \equiv   {\cal M} _\rho ( \lambda ){\cal M} _{ \rho } ( \lambda ' )\,,   \qquad   \Gamma \left( \lambda , \lambda ' \right) \equiv   {\cal M} _{ \Gamma  } (\lambda)  {\cal M} _{ \Gamma   } (\lambda')\,.
\end{equation} 
Here we define $ {\cal M}_{\rho}(\lambda)[{\cal M} _{ \Gamma } (\lambda) ] $ as the production [decay] amplitude of a top with spin $\lambda$. Furthermore, in this work we denote a particles 4-momenta with their symbol and $ m $ is the top quark mass. $\rho$ and $\Gamma$ are the spin density and decay matrices respectively. 

A complete treatment with general couplings requires the calculation of each element in these matrices, which takes into account the interference terms between spin up and spin down tops that are being produced. These interference terms can result in complicated expressions for the full cross-section in single top production (see for example, ref.~\cite{Rindani2011}). However, a powerful simplification can be made if the tops come out highly polarized. In this case, the outgoing particles are roughly pure spin states and interference terms are small. This requires a clever choice of spin quantization axis, instead of the helicity basis. The idea to consider a different spin basis has been successful in the ditop channel in maximizing polarization~\cite{shadmi1996,Parke1996}. Finding the appropriate basis to minimize correlations is non-trivial but has been done for the s-channel single top production using general couplings~\cite{Manzano2002,Hioki2003}. While deviating from the helicity basis makes the results more difficult to interpret it greatly simplifies the calculation. By choosing such a basis we are diagonalizing the spin density matrix, $\rho$. In that basis the results are also independent of the off-diagonal elements of the decay matrix, $\Gamma$, see Eq.~(\ref{eq:rhogam}). While here we consider only the s-channel, a polarization vector that diagonalizes the spin density matrix can be found for a variety of different channels \cite{Manzano2002,Espriu2002,Parke2000}. With this simplification the cross-section without hadronization can be approximated by~\cite{Tsai1973},
\begin{equation} 
d \sigma = \sigma _{\uparrow} \frac{ d \Gamma _{\uparrow}  }{ \Gamma } + \sigma _{\downarrow} \frac{ d \Gamma _{\downarrow} }{ \Gamma } \,,\label{eq:sigma}
\end{equation} 
where $\sigma_{\lambda}$ is the cross-section for producing a top of spin $\lambda$ and $d \Gamma_{\lambda}$ is the differential rate of the decay of such a top.

Hadronization, however, modifies this equation. So far we have assumed a spin up top stays spin up while a spin down stays spin down. Instead, we can think of the particles having an effective decay distribution given by 
\begin{align} 
& \frac{ d \Gamma _{\uparrow} }{  \Gamma } \;\longrightarrow\; \frac{ d \Gamma _{\uparrow} ^{ eff}} { \Gamma } = P (t) \frac{ d \Gamma _{\uparrow} }{ \Gamma } + [1- P (t) ] \frac{ d \Gamma _{\downarrow} }{ \Gamma }\,, \label{eq:Geff1}\\ 
& \frac{ d \Gamma _{\downarrow} }{ \Gamma } \;\longrightarrow\; \frac{ d \Gamma _{ \downarrow } ^{ eff}} { \Gamma } = [1- P (t) ]  \frac{ d \Gamma _{\uparrow} }{ \Gamma } + P (t) \frac{ d \Gamma _{\downarrow} }{ \Gamma } \,,
\label{eq:Geff2}
\end{align}  
where $P(t)$ is the probability a spin up (down) top will remain spin up (down) after a time $t$.

In principle the problem is solved. We can define a forward backward asymmetry from this expression that will be dependent on the hadronization. However, this is currently very abstract. In order to make progress we need four ingredients,
\begin{enumerate} 
\item A spin quantization axis such that interference terms between producing spin up and spin down top-like quarks are negligible. 
\item The top production cross-section into spin up and spin down, $ \sigma _{\uparrow} $ and $ \sigma_{ \downarrow } $.
\item The probability that a spin up top will stay spin up, $P(t)$.
\item The top decay distributions for a spin up and spin down top, $d \Gamma _{\uparrow} / \Gamma$ and $d \Gamma_{\downarrow} / \Gamma$.
\end{enumerate}

\section{$t$ Quark Production}
\begin{figure} 
\center
\begin{tikzpicture}
\large
\draw[thick,fb] (0,0) -- (1,1); \node[below left] at (0,0) {$\bar{d} $}; 
\draw[thick,f] (0,2) -- (1,1); \node[above left] at (0,2) {$ u $}; 
\draw[thick,v] (1,1) -- (3,1); \node[below] at (2,1) {$ W ^+\hspace{-0.2cm}\rightarrow $}; 
\draw[thick,fb] (3,1) -- ( 4,0); \node[below right] at (4,0) {$ \bar{b} $}; 
\draw[thick,f] (3,1) -- (4,2); \node[above right] at (4,2) {$ t $};
\end{tikzpicture}
\caption{The s-channel processes we consider for top-like quark production.}
\label{fig:schannel}
\end{figure}
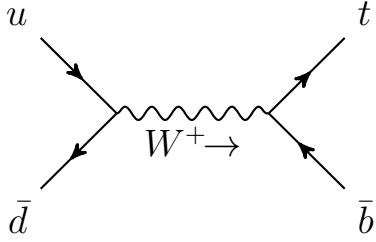

Consider top-like quark produced from the s-channel process shown in Fig.~\ref{fig:schannel}. 
In order to calculate the production cross-section of the hypothetical top quark we need to settle on a particular model. While we take the up-down vertex to be a Standard Model vertex, we consider a general vector coupling between the $t$ and $b$ 
\begin{equation} 
{\cal L}_{ int} = \frac{ g_W }{ \sqrt{2} }W_\mu\, t \,\bar b \, \gamma ^\mu \left( P_L f_L  + P_R f _R  \right)\,, \qquad P _{ R,L} \equiv ( 1 \pm \gamma ^5 )/2\,.
\label{eq:topvertex}
\end{equation} 
In the Standard Model (SM) for the real top we have $f_L = 1$ and $f_R = 0$. For our technique to work, the couplings need to be chiral, that is, $f_L \ne f_R$. Note that the above expression is not the most general possible coupling as we could have a derivative term vertex. While such terms could have been included, they add unnecessary complications without any more insight into the problem.

We are now ready to calculate the production cross-sections of the top-like quark. We find the amplitude for spin up and spin down decay separately using the spinor helicity method (for reviews see for example ref.~\cite{Nasuno1999,Huang2013,Andreev2001}). We introduce two massless momenta, 
\begin{equation} 
t_{1,2} \equiv \frac{ t \pm m \,s_t}{2} \, , \label{eq:t12def}
\end{equation} 
where $s_t$ is its spin quantization axis. We also use standard notations~\cite{Nasuno1999},
\begin{equation}
\ket{ p \pm } \equiv u _{ \pm } (p) = v _\mp (p) = P _{ R,L} u (p)\,, \qquad
\bra{ p \pm } \equiv \bar{u} _\pm (p) = \bar{v} _\mp (p) = \bar{ u} (p) P _{L,R}\,,
\end{equation}
with,
\begin{equation}
\left\langle p q \right\rangle \equiv \bar{u}  _- (p) u_+ (q) \,,\qquad \left[ p q \right] \equiv \bar{u}  _+ (p) u _- (q)  \, .
\end{equation}
We denote the amplitude for top production by $ {\cal M}_{ \lambda_{\bar{b}} \lambda_t } $ where $ \lambda_i $ denotes the spin of particle $i$ (the spin of the up and anti-down quarks is fixed by the chirality of the vertex). The amplitudes for this process in spinor helicity notation are given by: 
\begin{align} 
& {\cal M} _{ - + } = f _L 2\Delta \left[ u \bar{b} \right] \left\langle t _2 \bar{d} \right\rangle \frac{ \left[ t _1 t _2 \right] }{ m } ,\\ 
& {\cal M} _{ + + } = f _R 2 \Delta \left[ t _1 u \right] \left\langle \bar{d} \, \bar{b} \right\rangle ,\\[5pt]
& {\cal M} _{ - - } = f _L 2 \Delta \left[ u \bar{b} \right] \left\langle t _1 \bar{d} \right\rangle ,\\ 
& {\cal M} _{ + - }  = f _R 2 \Delta \left[ t _2 u \right] \left\langle \bar{d} \, \bar{b} \right\rangle \frac{ \left\langle t _1 t _2 \right\rangle }{ m },
\end{align} 
where we define 
\begin{equation}
\Delta \equiv  \frac{g _W ^2    V _{ ud}}{2 ( s - M _W ^2 + i M _W \Gamma_W) } \, .
\end{equation}
Here we denote $ M _W $ and $ \Gamma _W $ as the mass and width of the $W$ boson respectively. We have kept only the $W$ and top masses and sent the other masses to zero. 

The total amplitude for top production is 
\begin{equation} 
 \overline{ \left| {\cal M} \right| ^2}  =   \frac{ N _c }{ 2 ^2 }\left( \left| {\cal M} _{ - + } + {\cal M} _{ - - } \right| ^2 + \left| {\cal M} _{ + + } + {\cal M} _{ + - }\right| ^2 \right)  \,.
\end{equation} 
The interference terms are given by 
\begin{align} 
&  {\cal M} ^\ast _{ - - } {\cal M} _{ - + } = 4 \left|  \Delta \right|  ^2 \left| f _L \right| ^2 \left\langle \bar{b} u   \right\rangle \left[   u \,\bar{b} \right] \left[ \bar{d} \,t _1 \right]  \left\langle t _2 \,\bar{d} \right\rangle \frac{ \left[ t _1\, t _2 \right] }{ m } + h.c. \,,\label{flint}\\ 
& {\cal M} _{ + + } ^\ast {\cal M} _{ + - }  = 4 \left| \Delta \right| ^2 \left| f _R \right| ^2 \left[ \bar{b} \, \bar{d} \right] \left\langle \bar{d} \,\bar{b} \right\rangle \left\langle u\, t _1 \right\rangle \left[ t _2 \,u \right] \frac{ \left\langle t _1\, t _2 \right\rangle }{ m }  + h.c.\,.\label{frint}
\end{align} 
A general spin axis that minimizes the interference terms, proposed in \cite{Manzano2002}, for any values of $f_L$ and $f_R$ is given by
\begin{equation} 
  s ^\mu _t  =   A \bigg\{ \left| f _L \right| ^2 \Big[ ( u \cdot \bar{b} ) ( \bar{d} \cdot t ) t ^\mu - ( u \cdot \bar{b} ) \bar{d} \Big] + \left| f _R \right| ^2 \Big[ ( \bar{d} \cdot \bar{b} ) ( u \cdot t ) t ^\mu - ( \bar{d} \cdot \bar{b} ) m ^2 u ^\mu \Big] \bigg\}    \,, \label{eq:polvecgen}
\end{equation} 
where $A$ is a constant chosen such that $s^2=-1$. In the limit of small $\left| f_R \right|$, the vector is given by
\begin{equation} 
  s ^\mu_t  = \frac{1}{ m } \left( \frac{ m ^2 }{ \bar{d} \cdot t } \bar{d} ^\mu  - t ^\mu \right) \,.\label{eq:polarizationvec}
\end{equation} 
Choosing $s^\mu_t$ as the quantization axis sets $ \left[ \bar{d}\, t _1 \right] = 0 $ and eliminates the interference term in (\ref{flint}) that is proportional to $ \left| f _L \right| ^2 $, while it keeps the one in (\ref{frint}) that is proportional to $ \left|f_R \right|^2 $. Since we are assuming $ \left| f _R \right| $ to be small, the interference terms are small, justifying the choice of quantization axis. Furthermore, the interference terms are suppressed by $m/E$ and thus are unimportant at high energies. For these reasons we eliminate such terms from the discussion and keep only diagonal terms in the spin density matrix. In practice, the effects of the interference terms can always be estimated and included in the systematic uncertainties.

The total cross-sections are
\begin{align}
& \sigma _{\uparrow}= \frac{ s - m ^2  }{ 32\pi s ^2 } \Delta ^2 N _c  \bigg\{ \left| f _L \right| ^2 \frac{1}{3} \left( 2s ^2  - m ^2 s  - m ^4 \right)    \notag \\ 
&\hspace{4cm}  + \left| f _R \right| ^2 \left( \frac{ 2m ^2 s ^2  }{ s - m ^2 }  \log \frac{ s }{ m ^2 } - 2m ^2 s \right)  \bigg\}\,,   \label{eq:sigup}\\ 
& \sigma _{\downarrow} = \frac{ s - m ^2 }{ 32 \pi s ^2 } \Delta ^2 N _c  \left| f _R \right| ^2   \left\{ \frac{1}{3} \left( 2s ^2 + 5 m ^2 s - m ^4 \right)  - \frac{ 2 m ^2 s ^2 }{  s -m ^2 } \log \frac{ s }{ m ^2 } \right\} \,.
 \label{eq:sigdo}
\end{align} 
The sum of these two agrees with the results of ref.~\cite{Yuan2005}. Note that $\sigma_{\uparrow}$ and $\sigma_{\downarrow}$ depend of the choice of the quantization axis, but their sum does not.  In the limit that $\left| f_R \right|^2 \rightarrow 0$ the tops come out $ 100 \% $ polarized as in the SM~\cite{Mahlon2000}. This shows another advantage of using the spin axis direction chosen earlier, since the larger the polarization, the better our resolution. 

\begin{figure} 
\center
\includegraphics[width=\textwidth]{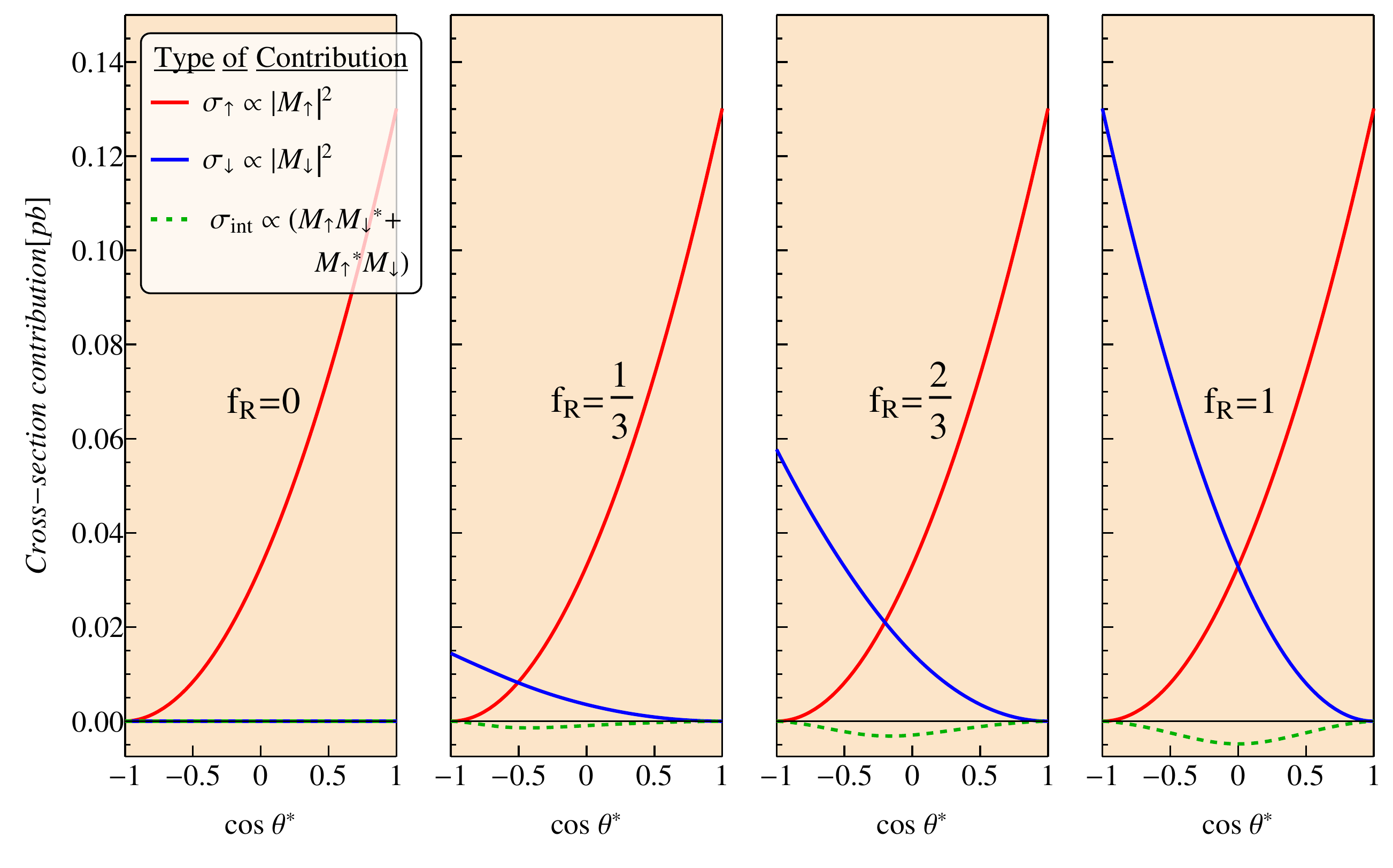}
\caption{(color online). The contribution to the top-like quark production cross-section of all the terms including the interference terms ($ \sigma _{ int} $), which are dropped in the final results. Here we use the general coupling polarization vector with $ f _L = 1 $, vary $ f _R $, and take $ M _W = 80.4\,\mbox{GeV}, \Gamma _W = 2.1\,\mbox{GeV},m = 300 \, \mbox{GeV}, $ and $ \sqrt{s} = 1\, \mbox{TeV} $. Clearly, the interference terms are much smaller than the dominant cross-section contributions.}
\label{fig:intercross}
\end{figure}

The different contributions to the total cross-section using the general spin vector given in Eq.~(\ref{eq:polvecgen}) with $f_L=1$ and different values of $f_R$ are shown in Fig.~\ref{fig:intercross}. The interference terms (shown in dashed green) are clearly much smaller then the dominant spin up top production cross-section, even for a non-negligible choice of $f_R$. This is a consequence of the choice of spin basis. As expected, in the  $f_R \rightarrow 0 $ these terms vanish.

\section{Decay Distribution}
To calculate the decay distribution we use the same model as above. We calculate the distributions for the spin up and spin down top decays separately. As before, we denote the amplitudes as ${\cal M}_{ \lambda  _{\bar{b}}  \lambda _t } $ where $ \lambda  _i $ denotes the spin of particle $i$. The amplitudes for the decays are:
\begin{align} 
  & {\cal M} _{ - - } = \frac{ g _W }{ \sqrt{2} }f _L \braket{ b - | {\slashed \epsilon}   | t _1 -  } \,,\\ 
  & {\cal M} _{ - + } = \frac{ g _W }{ \sqrt{2} }f _L \braket{ b - | {\slashed \epsilon} | t _2 - } \frac{ \left\langle t _2 t _1 \right\rangle }{ m }\,, \\ 
& {\cal M} _{ + - } = \frac{ g _W }{ \sqrt{2} }f _R \braket{ b + | {\slashed \epsilon} | t _2 + } \frac{ \left[ t _2 t _1 \right] }{ m } \,,\\ 
& {\cal M} _{ + + } = \frac{ g _W }{ \sqrt{2} }f _R \braket{ b + | {\slashed \epsilon} | t _1 + } \,,
\end{align}  
where we have neglected all masses but the top quark and $W$ boson masses. We denote ${\slashed \epsilon} \equiv \epsilon_\mu \gamma^\mu$ where $\epsilon _\mu $ is the polarization vector of the external $W$ boson.

The amplitude squared for the decay distribution for a spin-up and spin-down top quark are (we omit the color factor as its taken care of in the top production cross-sections)
\begin{equation} 
\overline{\left|  {\cal M} _{ \uparrow } \right| ^2 } \equiv  \left(   \left|  {\cal M}  _{ - + }  \right| ^2 + \left| {\cal M} _{ + + } \right| ^2 \right) \,, \qquad 
\overline{\left|  {\cal M} _{ \downarrow  } \right| ^2 } \equiv  \left(   \left|  {\cal M}  _{ - -  }  \right| ^2 + \left| {\cal M} _{ + -  } \right| ^2 \right) \,. 
\end{equation} 
These give the decay distributions,
\begin{equation} 
 \frac{1}{ \Gamma }\frac{ d \Gamma _{ \uparrow } }{ d \cos \theta ^\ast } = \frac{1}{2} \left( 1 -  \alpha \cos \theta ^\ast \right) \,, \qquad
 \frac{1}{ \Gamma }\frac{ d \Gamma _{ \downarrow } }{ d \cos \theta ^\ast } = \frac{1}{2} \left( 1 + \alpha \cos \theta ^\ast \right) \,,
\label{eq:Gud}
\end{equation} 
where 
\begin{equation} 
\alpha  \equiv \left(\frac{ \left| f _L \right| ^2 - \left| f _R \right| ^2} { \left| f _L \right| ^2 + \left| f _R \right| ^2 }\right)\times \left(\frac{ m ^2 - 2 M _W ^2 }{ m ^2 + M _W ^2 } \right)
\label{eq:alpha}
\end{equation}
and $\theta^\ast$ is the angle between the spin quantization axis of the $t$ quark, defined in Eq. (\ref{eq:polarizationvec}), and the direction of $b$ quark. This expression agrees with ref.~\cite{Najafabadi2006}. 
In the final result we do not have any cross terms proportional to $ \left|f _R \right| \left| f _L  \right|$. This is a consequence of taking the massless limit for the bottom quark. 

Note that $\theta^\ast$ is an observable. It is the angle between the axis that we quantize the spin of the top quark and the momenta of the $b$ quark in the top rest frame. Of course, the number of particles moving into different angles in the center of mass frame is independent of the choice of quantization axis. However, depending on this choice, the values of $ \sigma _{\uparrow} $ and $ \sigma _{\downarrow} $ change with $ \theta ^\ast $ to leave the center of mass variables invariant. By measuring the angles that the $b$ quark is emitted in the center of mass frame one can calculate $\theta^\ast$. For simplicity we express our results in terms of this angle. 

Eq.~(\ref{eq:Gud}) shows that a $1 - \alpha  \cos \theta^\ast$ distribution is characteristic of a spin up top decay while a $1 + \alpha \cos \theta^\ast$ distribution is characteristic of a spin down top decay. We will soon see the effect of having both such decays and the interaction between the two. We also see that we need a chiral theory. In a parity conserving theory we have $f_L = f_R$ and hence $\alpha = 0$. In this case we have the same distribution for a spin up and spin down top and they cannot be differentiated. Since we will use angular correlations to probe the lifetime of the decaying particles, the method fails in this case.

\section{The Effective Angular Distribution}
We are now in position to find the effective distribution. Using Eqs.~(\ref{eq:Geff1}), (\ref{eq:Geff2}), and (\ref{eq:Gud}) we write 
\begin{equation} 
\frac{ d \Gamma ^{ eff} _{ \uparrow \,,\,\downarrow } }{ d \cos \theta ^\ast } = \frac{1}{2} \left[ 1 \mp   ( 2 P (t)  - 1 ) \alpha \cos \theta ^\ast \right]\,. \label{eq:Geffdist}
\end{equation} 
The key dynamic quantity is $ 2P (t) - 1 $. Now note that the polarization of a set of tops given in Eq.~(\ref{eq:sz}) is related to the probability of a spin up top remaining spin up after a time $t$ by (if a top is measured spin up it contributions $ + 1/2 $ to the net angular momentum and if a top is measured spin down it contributes $ - 1/2 $ to the net angular momentum), 
\begin{equation} 
\frac{ \left\langle s \right\rangle (t) }{ \left\langle s \right\rangle (0) }  = \frac{1}{ 1/2} \left\{ \frac{1}{2} P (t) - \frac{1}{2} ( 1 - P (t) ) \right\}  = 2 P (t) - 1\,.
\end{equation}
So the dynamics are indeed controlled by the average polarization. 

\begin{figure} \center
  \includegraphics[width=10cm]{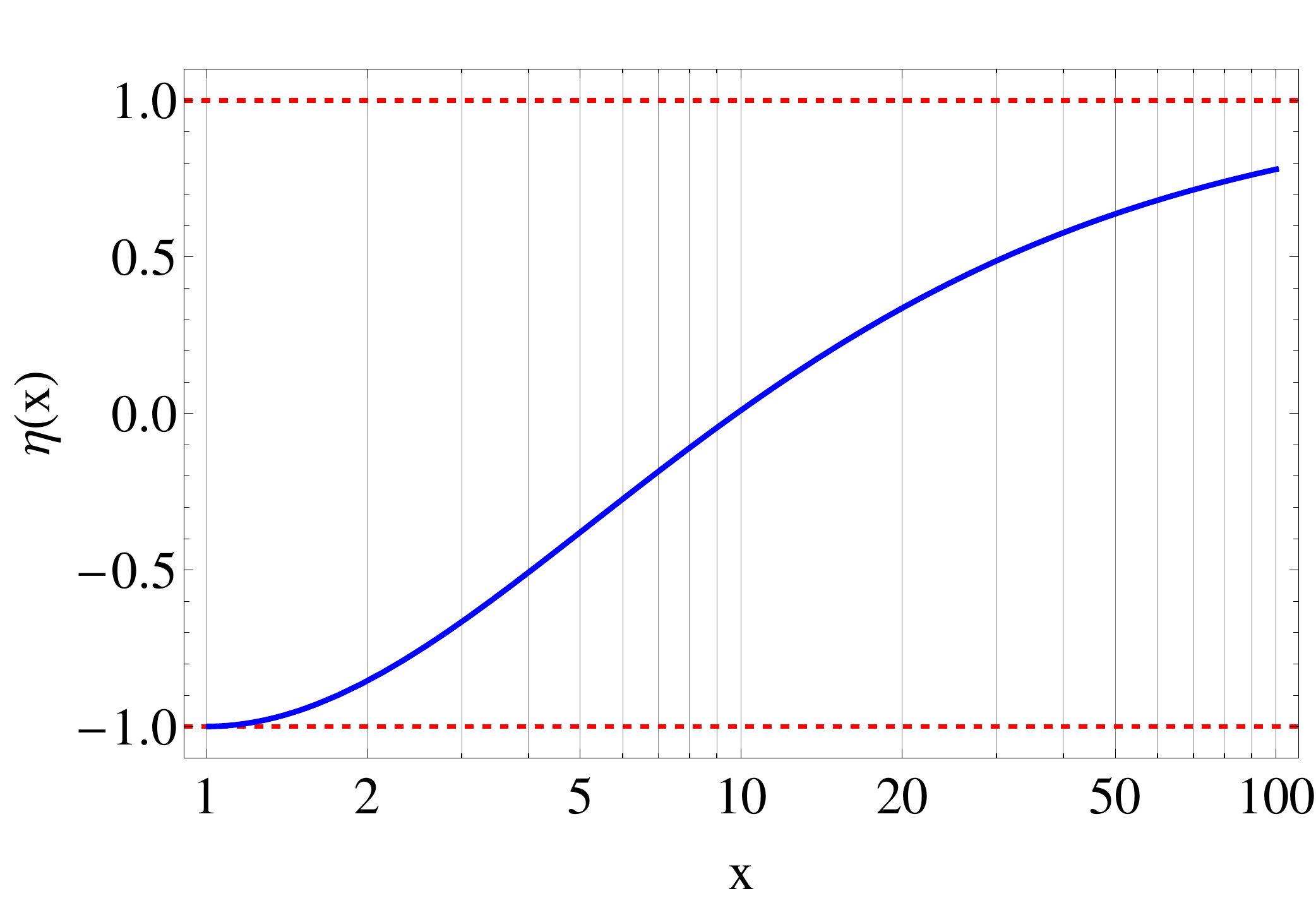}
\caption{$\eta$ as a function of center of mass energy. $\eta$ at low energies is $ - 1 $ but rapidly increases to $1 $ at larger energies.}
\label{fig:eta}
\end{figure} 

Top-like quarks come out with two opposing distributions. Depending on how many tops decay with spin up compared to the number coming out with spin down we have a $1 - \alpha \cos \theta^\ast$ or a $1 + \alpha \cos \theta^\ast$ dominated distribution. The average angular momentum of the set of tops, which oscillates as a function of time, determines the net distribution. 
Inserting the effective decay distributions into (\ref{eq:sigma}), we can write the distribution of $b$ quarks at different angles $\theta^\ast$ and times $t$,
\begin{align} 
\frac{ d \sigma (t) }{ d \cos \theta ^\ast  d t} & =  \Gamma e ^{ - \Gamma t } \frac{1}{2}\left\{  \sigma _{\uparrow} \left( 1 -   \frac{ \left\langle s \right\rangle (t) }{ \left\langle s  \right\rangle (0) } \alpha \cos \theta ^\ast \right) +  \sigma _{\downarrow } \left( 1 +   \frac{ \left\langle s \right\rangle (t) }{ \left\langle s  \right\rangle (0) } \alpha \cos \theta ^\ast \right) \right\}  \,, \nonumber \\ 
& = \Gamma e ^{ - \Gamma t } \left( \sigma _{\uparrow} + \sigma _{\downarrow} \right) \frac{1}{2} \left( 1 - \frac{ \sigma _{\uparrow} - \sigma _{\downarrow} }{ \sigma _{\uparrow} + \sigma _{\downarrow} }\alpha \frac{ \left\langle s \right\rangle (t) }{ \left\langle s \right\rangle (0) } \cos \theta ^\ast \right) \,,
\label{eq:dyncross}
\end{align} 
Here we have multiplied this dynamic cross-section by the probability density that a top quark lives until a time $t$. A more careful treatment of how to add the time dependence is given in Appendix \ref{sec:dyncross}. In the case of equal spin up and spin down production we lose the sensitivity to the lifetime. 

Thus far the discussion has been quite general and did not involve any specific channel. While the results do depend on the production cross-section, and in particular how polarized the production channel is, the details of the particular production channel have little importance. We now specialize to the s-channel by using Eqs.~(\ref{eq:sigup}) and (\ref{eq:sigdo}) and integrate over all time. This gives the distribution of outgoing $b$ quarks,
\begin{equation} 
\frac{1}{  \sigma  }\frac{ d   \sigma   }{ d \cos \theta ^\ast  } =  \frac{1}{2} \left\{ 1 -\left( \frac{ \left| f _L \right| ^2 - \eta ( s /m ^2 ) \left| f _R \right| ^2 }{ \left| f _L \right| ^2 + \left| f _R \right| ^2 }  \alpha \right)   r  \cos \theta ^\ast \right\} \,, \label{eq:distribution}
\end{equation} 
where 
\begin{equation} 
 \eta (x)  \equiv \frac{ \left(2 x ^3+  9 x^2 -12 x + 1\right) - 12x^2 \log ( x) }{ \left(x - 1 \right)^2 \left(2x + 1\right)}\,,
\end{equation} 
such that $-1\le \eta(x)\le 1$ but its exact value has little impact on the results. It is shown for different energies in Fig.~\ref{fig:eta}. Using this we can extract $r$ and hence the lifetime of the particle. Eq.~(\ref{eq:distribution}) is our main result. We have the angular dependence as a function of time. 

We are now in position to define a simple observable, a time integrated forward-backward asymmetry. We look at the total number of $b$ quarks at all times produced in different angles and define
\begin{equation}
a _{ fb}  \equiv \frac{ N _b ( \cos \theta ^\ast >0 ) - N _b ( \cos \theta ^\ast <0 ) }{ N _b ( \cos \theta ^\ast >0 ) + N _b ( \cos \theta ^\ast <0 ) } 
  =\left( \frac{ \sigma _{ \uparrow } - \sigma _{\downarrow}   }{ \sigma _{\uparrow} + \sigma _{\downarrow}  }\right)  \alpha \times r  \label{eq:afb0}\,.
\end{equation}
Specializing to the s-channel gives,
\begin{equation} 
a _{ fb} =  \left\{  \frac{ \left| f_L \right| ^2  - \eta  (s/m^2)\left| f_R \right| ^2    }{\left| f_L \right| ^2+\left| f_R \right| ^2 }  \,\,\frac{ \left| f _L \right|  ^2 - \left| f _R \right| ^2 }{ \left| f _L \right| ^2 + \left| f _R \right| ^2 } \,\,\frac{ 2 M _W ^2 - m ^2 }{ 2 M _W ^2 + m ^2 } \right\}   \times  r\,. \label{eq:afb}
\end{equation} 

In the limit of $\left| f _L \right|^2 \gg \left| f _R \right|^2 $, the dependence on $\eta$ drops out and we have the particularly simple result,
\begin{equation} 
a _{ fb } = \left\{ \frac{ 2 M _W ^2 - m ^2 }{ 2 M _W ^2 + m ^2 } \right\}  \times r \,. \label{eq:afbsimp}
\end{equation} 

\section{LHC, ILC, and Future Work}
The method discussed above requires collisions between $\bar{d}$ and $u$ to produce top quarks. This is not an ideal channel to produce top-like quarks at the LHC, which is dominated by pair productions from gluons, followed by the valence quarks, $u$ and $d$~\cite{Stirling2007}. However, at small momentum fractions, the number of sea quarks is significant and the $s$ channel discussed above is certainly feasible. Of course, if the top-like particle is heavy, single top-like production will be the dominate production mechanism. Moreover, also in the SM there are some simple and calculable alternatives to top-like production with significant polarization, including the $ t W ^- $ associated production and the dominant $W\!g$ fusion~\cite{Mahlon2000}. Since the $ t W ^- $ associated production suffers from either CKM suppression or requires a $b$ from the sea, it is suppressed. This makes the $W\!g$ fusion process, which uses a gluon and an up quark as the initial state, the most promising channel at the LHC with regards to measuring single top-like quark lifetimes in this way. 

The ILC is planned to have incoming beams with up to $~80\% $ polarization for one beam and $30\%$ for the other~\cite{Roeck2010}. While the discussion above focused on unpolarized initial states, having polarized incoming particles can increase the polarization of the top-like quarks~\cite{Vos2013}, yielding a better precision for the lifetime measurement technique we suggested. Of course, the channel discussed above is irrelevant for the ILC, which will collide electrons and positrons. Nevertheless, there exist both single top~\cite{Bouzas2011} and ditop production channels~\cite{Vos2013,Parke1996} which can be used instead. Although the references listed assume a SM coupling, any new physics coupling to top-like quarks that are chiral would yield similar results.

While in this work we study a top-like particle for simplicity, the most promising candidate for the use of this technique is the gluino, whose bound states include $ g \tilde{g} , q \bar{q} \tilde{g} , q \bar{q} \tilde{g} , $ and $ q \bar{q}  \tilde{g} $~\cite{Kundu1996}. The spectrum can be calculated using heavy gluino effective theory for both a Dirac~\cite{Okui2011} or Majorana~\cite{Kundu1996} gluino. 
Such studies and its quantitative impact on the lifetime measurement are left for future work.

In this paper we presented two observables that can be measured at a collider and are sensitive to the lifetime: the differential cross-section given in (\ref{eq:distribution}) and a forward-backward asymmetry, $ a _{ fb} $, derived from this cross-section given in (\ref{eq:afb}). $ a _{ fb} $ has the advantage of being particularly intuitive and emphasizes the importance of parity violation in our calculations. For low statistics this is the best measure to use. On the other hand, the forward backward asymmetry removes some of the information embedded in the cross-section. Eventually when more data is accumulated, fitting to the outgoing quark distribution would yield a more precise estimate of the lifetime.

In this paper we have focused on single top production and decay. However, the most common channel to study top production is through the ditop production channel. As before, there exists an appropriate choice of polarization vectors that minimizes the interference terms~\cite{Hioki2002}. Extending the lifetime measurement technique to this channel should be straightforward and likely more precise as typically the ditop channel contains many more events. A last requirement before particle lifetimes can be measured with the techniques presented here is to run simulations to test it.

To conclude, we implemented a new technique to measure lifetimes of top-like particles with lifetimes in the ``problematic region'' where current experiments cannot access. The complications associated with the calculations can be greatly reduced by choosing an appropriate spin basis that both maximizes polarization and eliminates extra terms. We arrive at a simple forward-backward asymmetry that is directly proportional to a quantity which characterizes the average angular momentum of top-like particles, $r$. This value is a function of the lifetime and its measurement allows direct access to the lifetime. 

\acknowledgments{We would like to thank Wee Hao Ng, and Yotam Soreq for looking over the manuscript and correspondences. JAD thank Randy Lewis for useful discussions. The work of JAD is supported in part by NSERC Grant PGSD3-438393-2013.
YG is a Weston Visiting Professor at the Weizmann Institute.
This work was partially supported by a grant from the Simons Foundation ($\#$267432 to Yuval Grossman).
The work of YG is supported is part by the U.S. National
Science Foundation through grant PHY-0757868 and
by the United States-Israel Binational Science
Foundation (BSF) under grant No.~2010221.}

\appendix 
\section{The Dynamic Cross-section}
\label{sec:dyncross}
Consider a sample of tops at a time $t$. Assume that initially the sample is completely polarized. The number of top quarks that are alive between $ t $ and $ t + \Delta t$ is 
\begin{equation} 
N (t) = N _0 \left( e ^{ - \Gamma t } - e ^{ - \Gamma ( t + \Delta t ) } \right) \,.\label{eq:Nt}
\end{equation} 
The initial number of particles is given by 
\begin{equation} 
 N _0 = {\cal L} \int d \sigma \,, \label{eq:N0}
\end{equation} 
where $ {\cal L} $ is the luminosity of the beam. The number of spin up tops which decayed between time $ t , t + \Delta t $ is 
\begin{equation} 
d N _{ decayed} = \left( {\cal L} \int d \sigma \right) \left( e ^{ - \Gamma t } - e ^{ - \Gamma ( t + \Delta t ) } \right) P  (t) \,,\label{eq:numtops}
\end{equation} 
where we denoted $ P  (t) $ as the probability that a spin up top will remain spin up at a time $t$ (neglecting its decay probability). 

The total number of spin up tops which decayed is given by 
\begin{equation} 
N _{ dec}= \sum _{ t}  \left( {\cal L} \int d \sigma \right) \left( \frac{ e ^{ - \Gamma t } - e ^{ - \Gamma ( t + \Delta t ) } }{ \Delta t }\right)  \Delta t P  (t) \,,\label{eq:totalN}
\end{equation} 
where the sum is over all possible times. Taking the $ \Delta t \rightarrow 0 $ limit we have 
\begin{equation} 
N _{ dec}= \int d t  \left( {\cal L} \int d \sigma \right)  \Gamma e ^{ - \Gamma t }  P  (t) \,.\label{eq:dtap0}
\end{equation} 
For every decaying top quark there is a corresponding $b$ quark that is emitted at some angle. Adding in a factor for the distribution of bottom quarks. The total number of $b$ quarks arriving at the detector is 
\begin{equation} 
N _ { t _{\uparrow}   \rightarrow t  _{\uparrow}  \rightarrow bW} = \int _0 ^{\infty} \bigg[ {\cal L} \int d \sigma \bigg] \bigg[ d t \Gamma e ^{ - \Gamma t } \bigg] \bigg[P  (t) \bigg] \left[ \int \frac{1}{ \Gamma  } \frac{ d \Gamma _{\uparrow}   }{ d \cos \theta ^\ast } d  \cos \theta ^\ast  \right] \,,\label{eq:Nb}
\end{equation}  
where in the subscript we have denoted the process we considered, producing spin up tops which remain spin up and decay. 

We now switch the expression into a differential cross-section,
\begin{equation} 
\frac{d  \sigma  _{ t _{\uparrow}   \rightarrow t  _{\uparrow}  \rightarrow bW }  }{ d \Omega d \cos \theta ^\ast d t }= \frac{  d \sigma ( u \bar{d} \rightarrow \bar{b} t _{\uparrow}  )  }{ d \Omega  }\bigg[  \Gamma e ^{ - \Gamma t } \bigg] \bigg[P  (t) \bigg] \left[ \frac{1}{ \Gamma _t } \frac{ d \Gamma _{\uparrow}   }{ d \cos \theta ^\ast }  \right]\,, \label{eq:cross1}
\end{equation}  
where $ d \Omega $ is the solid angle associated with top production. 
This calculation was done for one case where a spin up top was produced and stayed spin up when it decayed. Including all four cases, 
\begin{align} 
& u \bar{d} \rightarrow t _{\uparrow} \xrightarrow{had} t _{\uparrow}  \rightarrow b W\hspace{1cm} u \bar{d} \rightarrow  t _{\downarrow } \xrightarrow{had} t _{\uparrow }\rightarrow b W \nonumber \\ 
& u \bar{d} \rightarrow t _{\downarrow   } \xrightarrow{had} t _{ \downarrow   }  \rightarrow b W \hspace{1cm}  u \bar{d} \rightarrow t _{\uparrow  } \xrightarrow{had} t _{ \uparrow  }  \rightarrow b W
\end{align} 
and integrating over $ d \Omega  $ gives,
\begin{align} 
\frac{ d \sigma (t)  }{d \cos \theta ^\ast d t} &  = \Gamma e ^{ - \Gamma t }\bigg\{\bigg[ \sigma _{\uparrow}   P (t) + \sigma _{\downarrow}   (1- P (t)) \bigg] \frac{ d \Gamma _{ \uparrow }}{ d \cos \theta ^\ast } \notag \\ 
& \hspace{2cm}+  \bigg[ \sigma _{\downarrow}    P (t) + \sigma _{\downarrow}  (1- P (t)) \bigg] \frac{ d \Gamma _{ \downarrow  }}{ d \cos \theta ^\ast }\bigg\} \,,\label{eq:fineq}
\end{align} 
which is equivalent to (\ref{eq:dyncross}) once the differential distribution is added. 

\bibliographystyle{JHEP}
\bibliography{DrGr.bib}
\end{document}